\documentclass[pdftex,leqno,fleqn,12pt]{article}

\usepackage[latin1]{inputenc}   
\usepackage{xspace}

\usepackage{graphicx}           

\usepackage{latexsym}           
\usepackage{amsmath}
\usepackage{amssymb}
\usepackage{amsfonts}
\usepackage{subfigure}

\usepackage[sort&compress,numbers]{natbib}

\usepackage{algorithmic}
\usepackage{algorithm}

\newcommand{\DEL}{{\rm{Del}}}
\newcommand{\VOR}{{\rm{Vor}}}
\newcommand{\YAO}{{\rm{Yao}}}
\newcommand{\sdsr}{{\rm{SP-DT}}}

\newtheorem{theorem}{Theorem}[section]
\newtheorem{corollary}[theorem]{Corollary}
\newtheorem{lemma}[theorem]{Lemma}

\newtheorem{proposition}[theorem]{Proposition}

\newtheorem{definition}[theorem]{Definition}

\newtheorem{problem}[theorem]{Problem}

\newenvironment{proof}{{\textit Proof:} \rm}{\hfill $\square$ \medskip\\}
\newcommand{\old}[1]{{}}
\newcommand{\etal}{\textit{et al.}}

\begin{document}

\title{Spanners of Additively Weighted Point Sets\thanks{Research partially supported by NSERC, MRI, CFI, and MITACS.}}
\author{Prosenjit Bose \and Paz Carmi \and Mathieu Couture\\ {\small Scool of Computer Science, Carleton University, Ottawa, Canada}}


\maketitle

\begin{abstract} We study the problem of computing geometric spanners for
(additively) weighted point sets. A weighted point set is a set of pairs $(p,r)$ where $p$ is a
point in the plane and $r$ is a real number. The distance between two points $(p_i,r_i)$ and
$(p_j,r_j)$ is defined as $|p_ip_j|-r_i-r_j$. We show that in the case where all $r_i$ are positive
numbers and $|p_ip_j|\geq r_i+r_j$ for all $i,j$ (in which case the points can be seen as
non-intersecting disks in the plane), a variant of the Yao graph is a $(1+\epsilon)$-spanner that
has a linear number of edges. We also show that the Additively Weighted Delaunay graph (the
face-dual of the Additively Weighted Voronoi diagram) has constant spanning ratio. The straight
line embedding of the Additively Weighted Delaunay graph may not be a plane graph. 
We show how to compute a plane embedding that also has a
constant spanning ratio.  \end{abstract}

\newcommand{\UDG}{\rm UDG}
\newcommand{\bis}{\rm bis}

\newpage

\section{Introduction}

Let $G$ be a complete weighted graph where edges have positive weight. Given
two vertices $u,v$ of $G$, we denote by $\delta_G(u,v)$ the length of a
shortest path in $G$ between $u$ and $v$. A spanning subgraph $H$ of $G$ is
a \emph{$t$-spanner} of $G$ if $\delta_H(u,v)\leq t\delta_G(u,v)$ for all
pair of vertices $u$ and $v$. The smallest $t$ having this property is
called the \emph{spanning ratio} of the graph $H$ with respect to $G$. Thus,
a graph with spanning ratio $t$ approximates the $n \choose 2$ distances
between the vertices of $G$ within a factor of $t$. Let $P$ be a set of $n$
points in the plane. A \emph{geometric graph} with vertex set $P$ is an
undirected graph whose edges are line segments that are weighted by their
length.  The problem of constructing $t$-spanners of geometric graphs with
$O(n)$ edges for any given point set has been studied extensively; see the
book by Narasimhan and Smid~\cite{smid07} for an overview.

In this paper, we address the problem of computing geometric spanners with
additive constraints on the points. More precisely, we define a weighted
point set as a set of pairs $(p,r)$ where $p$ is a point in the plane and
$r$ is a real number. The distance between two points $(p_i,r_i)$ and
$(p_j,r_j)$ is defined as $|p_ip_j|-r_i-r_j$. The problem we address is to
compute a spanner of a complete graph on a weighted point set. To the best
of our knowledge, the problem of constructing a geometric spanner in this
context has not been previously addressed. We show how the Yao graph can be
adapted to compute a $(1+\epsilon)$-spanner in the case where all $r_i$ are
positive real numbers and $|p_ip_j|\geq r_i+r_j$ for all $i,j$ (in which
case the points can be seen as non-intersecting disks in the plane). In the
same case, we also how the Additively Weighted Delaunay graph (the face-dual
of the Additively Weighted Voronoi diagram) provides a plane spanner that
has the same spanning ratio as the Delaunay graph of a set of points.

\subsection{Motivations}

It has been claimed (see ~\cite{alzoubi03,schindelhauer04,schindelhauer07})
that geometric spanners can be used to address the link selection problem in
wireless networks. In most cases, however, two assumptions are made:
\begin{enumerate}
\item nodes can be represented as points in the plane and
\item the cost of routing a message is a function of the length of the links
that are successively used.
\end{enumerate}
However, these assumptions do
not always hold. For example, the first assumption does not hold in the case
of wide area mesh networks, where nodes are vast areas such as
villages~\cite{raman04}.  The second assumption does not take into account
the fact that some nodes may have higher energy resources or introduce more
delay than others. In such cases, an additional cost must be taken into
account for each node. The study of spanners of additively weighted
point sets is a first step in addressing some of these issues.

\subsection{Paper Organization}

The rest of the paper is divided as follows: In Section~\ref{section-disk-del-related}, we review
related work. In Section~\ref{section-disk-del-def}, we give a formal definition of our problem and
show that it is not solved by a straightforward extension of the Yao graph. However, in
Section~\ref{section-yao}, we show that a minor adjustment to the Yao graph allows to compute a
$(1+\epsilon)$-spanner. In Section~\ref{section-disk-del-quotient}, we develop some tools used in
Section~\ref{section-disk-del-spanning-ratio} to show that the Additively Weighted Delaunay graph
has a constant spanning ratio. We conclude in Section~\ref{section-disk-del-conclusion}.

\section{Related Work}\label{section-disk-del-related}

Well known examples of geometric $t$-spanners include the Yao
graph~\cite{yao82}, the $\theta$-graph~\cite{ruppert91}, the Delaunay
graph~\cite{keil92}, and the Well-Separated Pair Decomposition
(WSPD)~\cite{callahan95}. Let $\theta<\pi/4$ be an angle such that
$2\pi/\theta=k$, where $k$ is an integer. The Yao graph with angle $\theta$
is defined as follows. For every point $p$, partition the plane into $k$
cones $C_{p,1},\ldots,C_{p,k}$ of angle $\theta$ and apex $p$.  Then, there
is an oriented edge from $p$ to $q$ if and only if $q$ is the closest point
to $p$ in some cone $C_{p,i}$. The Yao graph is sometimes confused with the
$\theta$-graph, although they are different graphs. The first phase of the
construction of the $\theta$-graph using $k$ cones with angle $\theta$ and
apex $p$ is identical to the construction of the Yao graph. This may be the
root of the confusion. However, there is an oriented edge from $p$ to $q$ in
the $\theta$-graph if and only if $q$ has the shortest projection on the
bisector of the cone containing $q$. For Yao graphs~\cite{yao82}, the
spanning ratio is at most $1/(\cos\theta-\sin\theta)$ provided that
$\theta<\pi/4$, and for $\theta$-graphs, the spanning ratio is at
most~$1/(1-2\sin\frac{\theta}{2})$ provided that
$\theta<\pi/3$~\cite{ruppert91}.

Given a set of points in the plane, there is an edge between $p$ and $q$ in the
Delaunay graph if and only if there is an empty circle with $p$ and $q$ on its
boundary~\cite{keil92}.  If no four points are cocircular, then the Delaunay
graph is a uniquely defined near-triangulation. Otherwise, four or more
co-circular points may create crossings. In that case, removing edges that
cause crossings leads to \emph{a} Delaunay triangulation. Since our results
hold for any Delaunay triangulation, when we refer to \emph{the} Delaunay
triangulation in the case of co-circular points, we mean \emph{any} Delaunay
triangulation. \citet{dobkin90} showed that the Delaunay triangulation has a
spanning ratio of at most $\frac{1+\sqrt{5}}{2}\pi\approx 5.08$. This result
was improved by~\citet{keil92}, who showed that the spanning ratio of the
Delaunay triangulation is at most $2\pi/(3\cos(\pi/6))\approx 2.42$.  Later,
\citet{bose04} showed that the Delaunay triangulation is also a strong
$t$-spanner for the same constant $t=2\pi/(3\cos(\pi/6))$.  Although the exact
spanning ratio of the Delaunay triangulation is unknown, it is conjectured that
the spanning ratio is $\pi/2$. For the remainder of this paper, we will refer
to the spanning ratio of the Delaunay triangulation as the spanning ratio of
the {\em standard} Delaunay triangulation and denote it as \sdsr.

\begin{figure}
\centering
\includegraphics{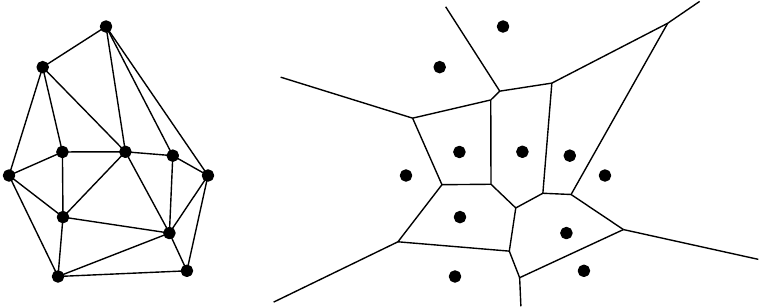}
\caption{The Delaunay graph and its dual: the Voronoi diagram.}\label{fig-vor-del}
\end{figure}

The \emph{Voronoi diagram}~\cite{deberg97} of a finite set of points $P$ is
a partition of the plane into $|P|$ regions such that each region contains
exactly those points having the same nearest neighbor in $P$. The points in
$P$ are also called \emph{sites}.  It is well known that the Voronoi diagram
of a set of points is the face dual of the Delaunay graph of that set of
points~\cite{deberg97}, i.e. two points have adjacent Voronoi regions if and
only if they share an edge in the Delaunay graph (see
Figure~\ref{fig-vor-del}).

Let $s>0$ be a real number. Two set of points $A$ and $B$ in $\mathbb{R}^d$ are
\emph{well-separated with respect to $s$} if there exists two $d$-dimensional
balls $C_A$ and $C_B$ of same radius $r$ respectively containing the bounding
boxes of $A$ and $B$ such that the distance between $C_A$ and $C_B$ is greater
than or equal to $s\times r$. The distance between $C_A$ and $C_B$ is defined
as the distance between their centers minus $2\times r$. A \emph{Well-Separated
Pair Decomposition with separation ratio $s$ of a set of points
$P$}~\cite{callahan95,smid07} is a set of unordered pairs
$\{\{A_1,B_1\},\ldots,\{A_m,B_m\}\}$ of subsets of $P$ that are well-separated
with respect to $s$ with the additional property that for every two points
$p,q\in P$ there is exactly one pair $\{A_i,B_i\}$ such that $p\in A_i$ and
$q\in B_i$. \citet{callahan95} showed that for $s>4$, every point set admits a
WSPD with separation ratio $s$ of $O(n)$ size that can be computed in $O(n\log
n)$ time. Choosing one edge per pair allows to construct a $t$-spanner that has
$O(n)$ size with $t=(s+4)/(s-4)$.

Our work falls in the context of computing spanners for geometric graphs other than the complete
Euclidean graph. Typically, variations of the spanner problem arise by either changing the distance
function or removing edges from the complete graph. For example, for a set $P$ of points in the
plane and a set $C$ of non-intersecting line segments whose endpoints are in $P$, the
\emph{visibility graph} of $P$ with respect to $C$ is the geometric graph with vertex set $P$ and
there is an edge $(pq)$ if and only if the segment $\overline{pq}$ is in $C$ or it does not cross
any segment in $C$ (in that case, $p$ and $q$ are said to be \emph{visible}). A spanner of the
visibility graph should then approximate Euclidean distances for every pair of points that are
visible from each other. The constrained Delaunay triangulation (a variation of the Delaunay
triangulation) is a $2.42$-spanner of the visibility graph~\cite{ioannis01, klein06, bose06}.

Unit disk graphs~\cite{hale80,johnson90} received a lot of attention from the wireless community. A
\emph{unit disk graph} is a graph whose nodes are points in the plane and edges join two points
whose distance is at most one unit. It is well-known that intersecting a unit disk graph with the
Delaunay or the Yao graph of the points provides a $t$-spanner of the unit disk
graph~\cite{bose04}, where the constant $t$ is the same as the one of the original graph. However,
this simple strategy does not work with all spanners. In particular, it does not work with the
$\theta$-graph~\cite{couture07c}. Unit disk graphs can be seen as intersection graphs of disks of
same radius in the plane. The general problem of computing spanners for geometric intersection
graphs has been studied by~\citet{furer07}.

Another graph that has been looked at is the \emph{complete $k$-partite Euclidean graph}. In that
case, points are assigned a unique color (which may be thought of as a positive integer) between 1
and $k$, and there is an edge between two points if and only if they are assigned different colors.
Bose \etal~\cite{couture07bispanReport} showed that the WSPD can be adapted to compute a
$t$-spanner of that graph that has $O(n)$ edges for arbitrary values of $t$ strictly greater than
5.

For spanners of arbitrary geometric graphs, much less is known. Alth{\"o}fer \emph{et
al.}~\cite{addjs-sswg-93} have shown that for any $t>1$, every weighted graph $G$ with $n$ vertices
contains a subgraph with $O(n^{1+2/(t-1)})$ edges, which is a $t$-spanner of $G$. Observe that this
result holds for any weighted graph; in particular, it is valid for any geometric graph. For
geometric graphs, a lower bound was given by Gudmundsson and Smid~\cite{gs-osogg-06}: They proved
that for every real number $t$ with $1 < t < \frac{1}{4} \log n$, there exists a geometric graph
$H$ with $n$ vertices, such that every $t$-spanner of $H$ contains $\Omega( n^{1 + 1/t} )$ edges.
Thus, if we are looking for spanners with $O(n)$ edges of arbitrary geometric graphs, then the best
spanning ratio we can obtain is $\Theta(\log n)$.

In the literature, spanners that use a distance other than the Euclidean distance have also been
proposed. For example, in a \emph{power}
spanner~\cite{aurenhammer87,li01,grunewald02,schindelhauer04}, the distance used to measure the
length of an edge is the square of the Euclidean distance between its two end points. This models
the fact that in wireless networks, the amount of energy needed to send a packet is proportional to
a power (not necessarily the square, however) of the Euclidean distance between the sender and
receiver~\cite{pahlavan95}. When reducing the latency is more important than reducing the amount of
energy being used, a \emph{hop} spanner~\cite{alzoubi03}, which gives an equal weight to every
edge, can be used.

In this paper, the Additively Weighted Voronoi diagram (AW-Voronoi diagram) is
of particular interest.
Its definition is similar to that of the (standard) Voronoi diagram, except
that each site $p_i$ is assigned a weight which is a real number $r_i$.
Weights are used to define a weighted distance. More detail about how the
weighted distance is used to define the AW-Voronoi diagram is given in
Section~\ref{section-disk-del-spanning-ratio}. The Additively Weighted Delaunay
graph (AW-Delaunay graph) is defined as the face-dual of the AW-Voronoi
diagram. Properties of the AW-Voronoi diagram and its dual have been studied by
Lee and Drysdale~\cite{drysdale81}, who showed how to compute it in $O(n\log^2
n)$ time. Later on, Fortune~\cite{fortune87} showed how to compute it in
$O(n\log n)$ time. The AW-Voronoi diagram may have empty cells. For this
reason, one would hope that it is possible to design an algorithm whose running
time gets better as the number of empty cells increases. Karavelas and
Yvinec~\cite{karavelas02} provided an $O(nT(h)+h\log h)$ time algorithm to
compute the AW-Voronoi diagram where $h$ is the number of non-empty cells and
$T(h)$ is the time to locate the nearest neighbor of a query point within a set
of $h$ points.  Experimental results suggested an $O(n\log h)$ behavior. In 3D,
the complexity of the (Additively Weighted) Voronoi diagram is
$\Theta(n^2)$~\cite{klee80}. Aurenhammer~\cite{aurenhammer87} showed how to
compute it in time $O(n^2)$ using Power Voronoi diagrams. Will~\cite{will98}
gave an $O(n^2\log n)$ time algorithm with experimental results suggesting an
$O(n\log^2 n)$ time behavior in the expected case. Kim \etal~\cite{kim05}
showed how to obtain a running time of $O(nm)$, where $m$ is the number of
edges.

\section{Definitions and Notation}\label{section-disk-del-def}

\begin{definition}
A set $P=\{(p_1,r_1),\ldots,(p_n,r_n)\}$ of ordered pairs, where each $p_i$ is a point in the plane
and each $r_i$ is a real number, is called a \emph{weighted point set}. The notation $p_i\in P$
means that there exists an ordered pair $(p_i,r_i)$ such that $(p_i,r_i)\in P$. The \emph{additive
distance} from a point $p\not\in P$ in the plane to a point $p_i\in P$, noted $d(p,p_i)$, is
defined as $|pp_i|-r_i$, where $|pp_i|$ is the Euclidean distance from $p$ to $p_i$. The \emph{additive distance}
between two points $p_i,p_j\in P$, noted $d(p_i,p_j)$, is defined as $|p_ip_j|-r_i-r_j$, where $|p_ip_j|$
is the Euclidean distance from $p_i$ to $p_j$.
\end{definition}
The problem we address in this paper is the following:
\begin{problem} Let $P$ be a weighted point set and let $K(P)$ be
the complete weighted graph with vertex set $P$ and edges weighted
by the additive distance between their endpoints. Compute a
$t$-spanner with $O(n)$ edges of $K(P)$ for a fixed constant $t>1$.
\end{problem}

Notice that in the case where all $r_i$ are positive numbers, the pairs
$(p_i,r_i)$ can be viewed as disks $D_i$ in the plane. If, for all $i,j$ we
also have $d(p_i,p_j)\geq 0$, then the disks are disjoint. In that case, the
distance $d(D_i,D_j)=d(p_i,p_j)=|p_ip_j|-r_i-r_j$ is also equal to
$\min\{|q_iq_j|:q_i\in D_i$ and $q_j\in D_j\}$, where the notation $q_i\in D_i$
means $|p_iq_i|\leq r_i$.  To compute a spanner of an additively weighted point
set is then equivalent to computing a spanner of a set of disks in the plane.
\textbf{From now to the end of this paper, it is assumed that all $r_i$ are
positive numbers and $d(p_i,p_j)\geq 0$ for all $i,j$.} If $\mathcal{D}$ is a
set of disks in the plane, then a \emph{spanner} of $\mathcal{D}$ is a spanner
of the complete graph whose vertex set is $\mathcal{D}$ and whose edges
$(D_i,D_j)$ are given weights equal to $d(D_i,D_j)$.

\begin{figure}
\centering\includegraphics{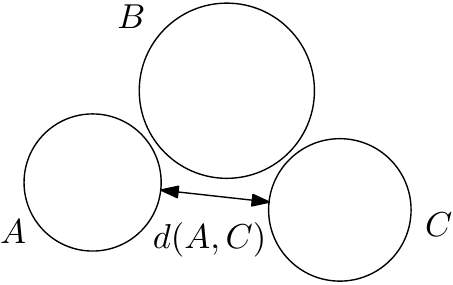}\caption{The additively weighted distance is not a
metric.}\label{fig-nometric}
\end{figure}

Notice also that the additive distance may not be a metric since the triangle inequality does not
necessarily hold (see Figure~\ref{fig-nometric}). Although this may seem counter-intuitive, this
makes sense in some networks, since a direct communication is not always easier than routing
through a common neighbor. For example, in wireless networks, the amount of energy that is needed
to transmit a message is a power of the Euclidean distance between the sender and the receiver.
Therefore, using several small hops can be more energy efficient that a direct communication over
one long-distance link.

\begin{figure}[htb]
\begin{center}
\includegraphics[scale=0.95]{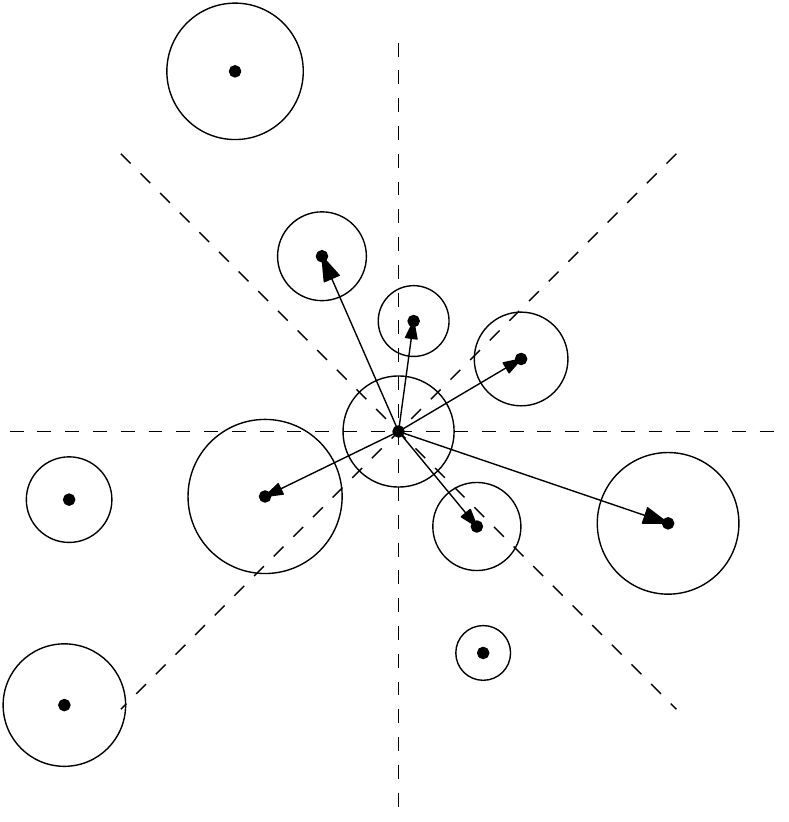} \caption{A straightforward generalization of the Yao graph.}\label{fig-yao-disks}
\end{center}
\end{figure}

\begin{figure}[htb]
\begin{center}
\includegraphics[scale = 0.85]{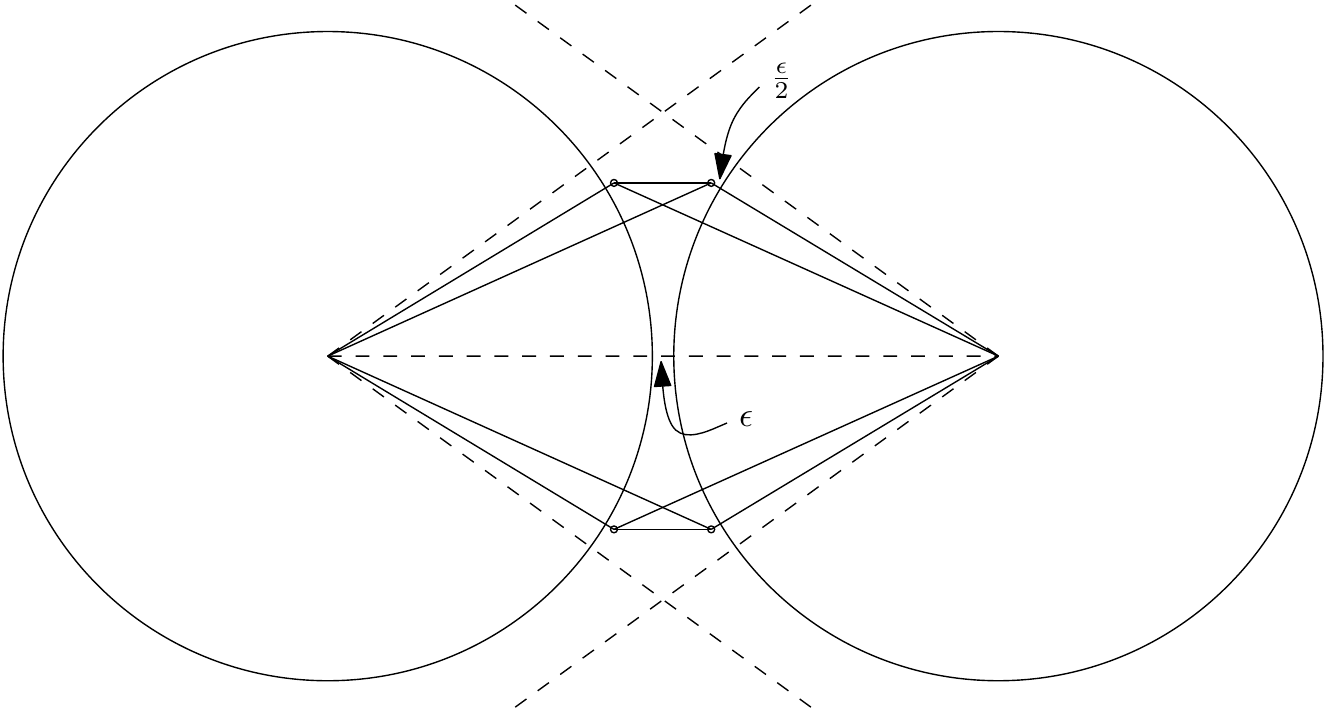}\caption{The straightforward generalization of the Yao graph
does not have constant spanning ratio.}\label{fig-yao-disks-not-spanner}
\end{center}
\end{figure}

Figure~\ref{fig-yao-disks} shows how the Yao graph can be generalized using the
additive distance: every node keeps an outgoing edge with the closest disk that
intersects each cone. However, this graph is not a spanner.
Figure~\ref{fig-yao-disks-not-spanner} shows how to construct an example with
four disks that has an arbitrarily large spanning ratio. Nonetheless, in
Section~\ref{section-yao}, we see that a minor adjustment to the Yao graph can
be made in order to compute a $(1+\epsilon)$-spanner of a set of disjoint disks
that has $O(n)$ edges.

The Delaunay graph in the additively weighted setting is computable in time
$O(n\log n)$~\cite{fortune87}. To the best of our knowledge, its spanning
properties have not been previously studied. In the two next sections, we show
that it is a spanner and that its spanning ratio is \sdsr\ (i.e the same as
that of the standard Delaunay graph). Finally, we show that although the
additively weighted Delaunay graph is not necessarily plane,  it contains a
plane subgraph that is a spanner with the same spanning ratio.
\section{The Additively Weighted Yao Graph}\label{section-yao}
\begin{figure} \centering\includegraphics{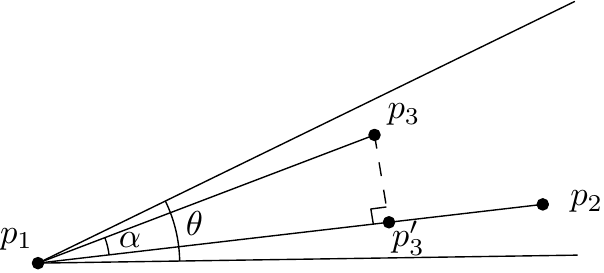}\caption{Illustration of
the proof of Lemma~\ref{lemma-yao}.}\label{fig-lemma-yao} \end{figure}
As we saw in the previous section, a straightforward generalization of the
Yao graph fails to provide a graph with bounded spanning ratio. In this section,
we show how a few subtle modifications to the construction, provide
an approach to build a $(1+\epsilon)$-spanner. We define the modified
Yao construction below.
\begin{definition} Let $\mathcal{D}$ be a finite set of disjoint disks and
$\theta\leq 0.228$ be an angle such that $2\pi/\theta=k$, where $k$ is an
integer. The $\YAO(\theta,\mathcal{D})$ graph is defined as follows. For every
disk $D=(p,r)$, partition the plane into $k$ cones $C_{p,1},\ldots,C_{p,k}$ of
angle $\theta$ and apex $p$. A disk \emph{blocks} a cone $C_{p,i}$ provided
that the disk intersects both rays of $C_{p,i}$. Let $F\in \mathcal{D}$ be a
disk different from $D$ with center in $C_{p,j}$. Add an edge from $D$
to $F$ in $\YAO(\theta,\mathcal{D})$ if and only if one of the two following
conditions is met: \begin{enumerate} \item among all blocking disks that have
their center in $C_{p,j}$, $F$ is the one that is the closest to $D$; \item
among all disks that have their center in $C_{p,j}$ and are at a distance of at
least $r$ to $D$, $F$ is the one that is the closest to $D$.  \end{enumerate}
\end{definition}
Notice that there are two main changes. Within each cone, we now add
potentially two edges as opposed to only one edge in the case of unweighted
points. Next, in the second condition to add an edge, we do not add an edge to
the closest disk within a cone but to the closest disk whose distance is at
least $r$ from the disk centered at the apex with radius $r$. We now prove that
these two modifications imply that the resulting graph is a $(1 +
\epsilon)$-spanner.
\begin{lemma}\label{lemma-yao} Let $p_1,p_2,p_3$ such that the angle $\angle
p_3p_1p_2=\alpha\leq\theta<\pi/4$ and $|p_1p_3|\leq |p_1p_2|$. Then
$|p_2p_3|\leq |p_1p_2|-(\cos\theta-\sin\theta)|p_1p_3|$.  \end{lemma}
\begin{proof} Let $p_3'$ be the projection of $p_3$ on the line through $p_1$
and $p_2$ (see Figure~\ref{fig-lemma-yao}). Then \begin{eqnarray*} |p_2p_3| &
\leq & |p_2p_3'|+|p_3'p_3|\\ & = & |p_1p_2| - |p_1p_3'| + |p_3'p_3|\\ & = &
|p_1p_2| - |p_1p_3|(\cos\alpha-\sin\alpha)\\ &\leq& |p_1p_2| -
|p_1p_3|(\cos\theta-\sin\theta) \end{eqnarray*} \end{proof}


%
\begin{theorem}\label{thm-add-yao} Let $\mathcal{D}$ be a finite set of
disjoint disks and $\theta\leq 0.228$.  Then $Y(\theta,\mathcal{D})$ is a
$t$-spanner of $\mathcal{D}$, where $t=1/(\cos 2\theta-\sin
2\theta-2\sin(\theta/2))$.  \end{theorem} \begin{proof} We proceed by induction
on the rank of the weighted distances between the pairs of disks $D_1$ and
$D_2$.

\textbf{Base case:} The disks $D_1$ and $D_2$ form a closest pair. In that case, the edge
$(D_1,D_2)$ is in $\YAO(\theta,\mathcal{D})$.

\textbf{Induction case:} Let $D_1=(p_1,r_1)$ and $D_2=(p_2,r_2)$. Without loss
of generality, $r_1\leq r_2$. If the edge $(D_1,D_2)$ is in
$\YAO(\theta,\mathcal{D})$, then there is nothing to prove.  Otherwise, there
are two cases to consider depending on whether or not the shortest path from
$D_1$ to $D_2$ in the complete graph on $\mathcal{D}$ is the edge $(D_1,D_2)$.
If the shortest path is not the edge $(D_1,D_2)$, then all edges on the
shortest path must have length less than $d(D_1,D_2)$. By applying the
induction hypothesis on each of those edges, we conclude that the distance from
$D_1$ to $D_2$ in $\YAO(\theta,\mathcal{D})$ is at most $t$ times the length of
the shortest path $D_1$ to $D_2$ in the complete graph on $\mathcal{D}$, as
required.

\begin{figure}
\centering\includegraphics{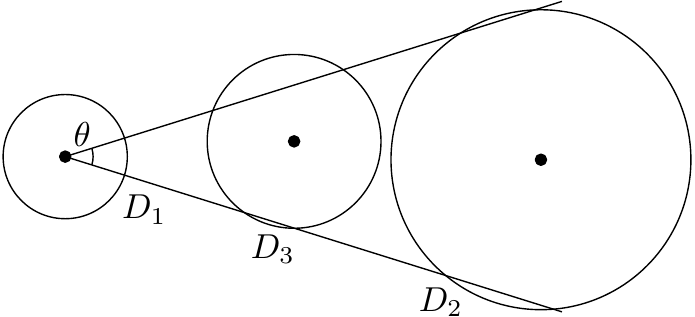}\caption{If $D_2$ blocks the cone but the edge $(D_1,D_2)$ is not in $\YAO(\theta,\mathcal{D})$, then there exists $D_3$ such that $d(D_1,D_3)+d(D_3,D_2)<d(D_1,D_2)$.}\label{fig-no-block}
\end{figure}

We now consider the case when the edge $(D_1,D_2)$:
\begin{enumerate}
\item is not in $\YAO(\theta,\mathcal{D})$ and
\item is the shortest path from $D_1$ to $D_2$ in the complete graph.
\end{enumerate}

Observe that the conjunction of those two facts imply that the disk $D_2$ does
not block the cone whose apex is $p_1$ and contains $p_2$: If $D_2$ was
blocking the cone, then since $(D_1,D_2)$ is not an edge in
$\YAO(\theta,\mathcal{D})$, there must be a disk $D_3$ that is also blocking
the cone and is closer to $D_1$ than $D_2$. However, this implies that the
shortest path from $D_1$ to $D_2$ in the complete graph is not the edge
$(D_1,D_2)$ (see Figure~\ref{fig-no-block}).

The conjunction of the three following facts:
\begin{enumerate}
\item $r_1\leq r_2$;
\item $\theta \leq 0.228 < \sin^{-1}(1/3)$ and
\item $D_2$ does not block the cone,
\end{enumerate}
imply that $d(D_1,D_2)>r_1$. Since $(D_1,D_2)$ is not an edge, there another
disk whose distance is at least $r$ that is closer to $D_1$. Let
$D_3=(p_3,r_3)$ be the closest disk to $D_1$ such that $p_3$ is in the same
$\theta$-cone with apex at $p_1$ as $p_2$ and $d(D_1,D_3)\geq r_1$. By
definition, the edge $(D_1,D_3)$ is in $\YAO(\theta,\mathcal{D})$.  Observe
that $d(D_2,D_3)<d(D_1,D_2)$. To see this, let $a:=d(D_1,D_2)-r_1$. We have
that $$d(D_2,D_3) \leq a + 4r_1\sin(\theta/2)\leq a + 4r_1\sin(0.114) < a+r_1 =
d(D_1,D_2).$$
\begin{figure} \centering\includegraphics{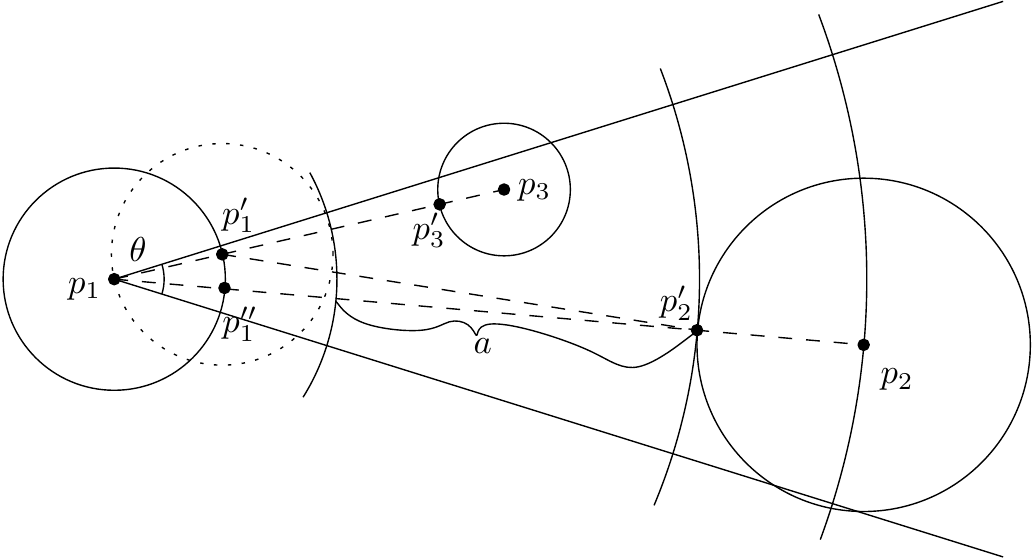}\caption{Illustration
of the proof of Theorem~\ref{thm-add-yao}.}\label{fig-add-yao} \end{figure}
Let $p_1'$ be the point of $D_1$ that is the closest to $D_3$, $p_1''$ be the
point of $D_1$ that is the closest to $D_2$, $p_2'$ be the point of $D_2$ that
is the closest to $D_1$, and $p_3'$ be the point of $D_3$ that is the closest
to $D_1$ (see Figure~\ref{fig-add-yao}). Notice that $|p_1'p_3'|\leq
|p_1'p_2'|$ and that since $d(D_1,D_2)\geq d(D_1,D_3) \geq r_1$, then the angle
$\angle p_2'p_1'p_3'$ is at most $2\theta<\pi/4$.  Therefore, we can apply
Lemma~\ref{lemma-yao} to conclude that $$|p_2'p_3'|\leq |p_1'p_2'|-(\cos
2\theta-\sin 2\theta)|p_1'p_3'|,$$ which implies that $$d(D_2,D_3)\leq
d(D_1,D_2)+|p_1'p_1''|-(\cos 2\theta-\sin 2\theta)d(D_1,D_3).$$ Also, since
$|p_1'p_1''|\leq 2\sin(\theta/2)r_1\leq 2\sin(\theta/2)d(D_1,D_3)$, we have
$$d(D_2,D_3)\leq d(D_1,D_2)-(\cos 2\theta-\sin
2\theta-2\sin(\theta/2))d(D_1,D_3).$$

Finally, since $d(D_2,D_3)<d(D_1,D_2)$, the induction hypothesis tells us that
$\YAO(\theta,\mathcal{D})$ contains a path from $D_2$ to $D_3$ whose length is at most
$td(D_2,D_3)$. This means that the distance from $D_1$ to $D_2$ in $\YAO(\theta,\mathcal{D})$ is at
most
$$d(D_1,D_3)+td(D_2,D_3)\leq d(D_1,D_3) + t(d(D_1,D_2)-\frac{1}{t}d(D_1,D_3))=td(D_1,D_2).$$
The value 0.228 is an upper bound on the values of $\theta$ such that $t>0$.
\end{proof}

\begin{corollary} For any $\epsilon>0$ and any set $\mathcal{D}$ of $n$ disjoint disks, it is
possible to compute a $(1+\epsilon)$-spanner of $\mathcal{D}$ that has $O(n)$ edges.
\end{corollary}
\begin{proof} The bound on the number of edges comes from the fact that each cone contains at most
two edges, and the stretch factor of $1+\epsilon$ comes from the fact that
$\lim\limits_{\theta\rightarrow 0}1/(\cos 2\theta-\sin 2\theta-2\sin(\theta/2))=1$.
\end{proof}

\section{Quotient Graphs and Quotient Spanners}\label{section-disk-del-quotient}

The main idea in the remainder of this paper is the following: we show how to compute a set of points
from each $D_i$ such that the (standard) Delaunay graph of those points is \emph{equivalent} to the
Additively Weighted Delaunay graph. By choosing the appropriate equivalence relation as well as the
appropriate point set, we can then show that the spanning ratio of the Additively Weighted Delaunay
graph is bounded by the spanning ratio of the standard Delaunay graph. The reduction of one graph
to another is done by means of a quotient:

\begin{definition} Let $P_1$ and $P_2$ be non-empty sets of points in the plane.
The \emph{distance} between $P_1$ and $P_2$, denoted by $|P_1P_2|$, is defined as the minimum $|p_1p_2|$
over all pairs of points such that $p_1\in P_1$ and $p_2\in P_2$.
\end{definition}

\begin{definition}Let $G=(V,E)$ be a geometric graph
and $\mathcal{V}$ be a partition of $V$. The \emph{quotient graph} of $G$ by $\mathcal{V}$, denoted
$G/\mathcal{V}$, is the graph having $\mathcal{V}$ as vertices and there is an edge $(U,W)$ (where
$U$ and $W$ are in $\mathcal{V}$) if and only if there exists an edge $(u,w)\in E$ with $u\in U$
and $w\in W$. The weight of the edge $(U,W)$ is equal to $|UW|$.
\end{definition}

If $P$ is a (non-weighted) point set and $\mathcal{P}$ is a partition of $P$, then the notation
$P/\mathcal{P}$ designates the quotient of the complete Euclidean graph on $P$ by $\mathcal{P}$. If
$\mathcal{S}$ is a set of pairwise disjoint sets of points in the plane such that $P\subseteq
\bigcup\mathcal{S}$, then the notation $P/\mathcal{S}$ designates the quotient of the complete
Euclidean graph on $P$ by the partition of $P$ induced by $\mathcal{S}$.

\begin{figure}
\centering
\includegraphics{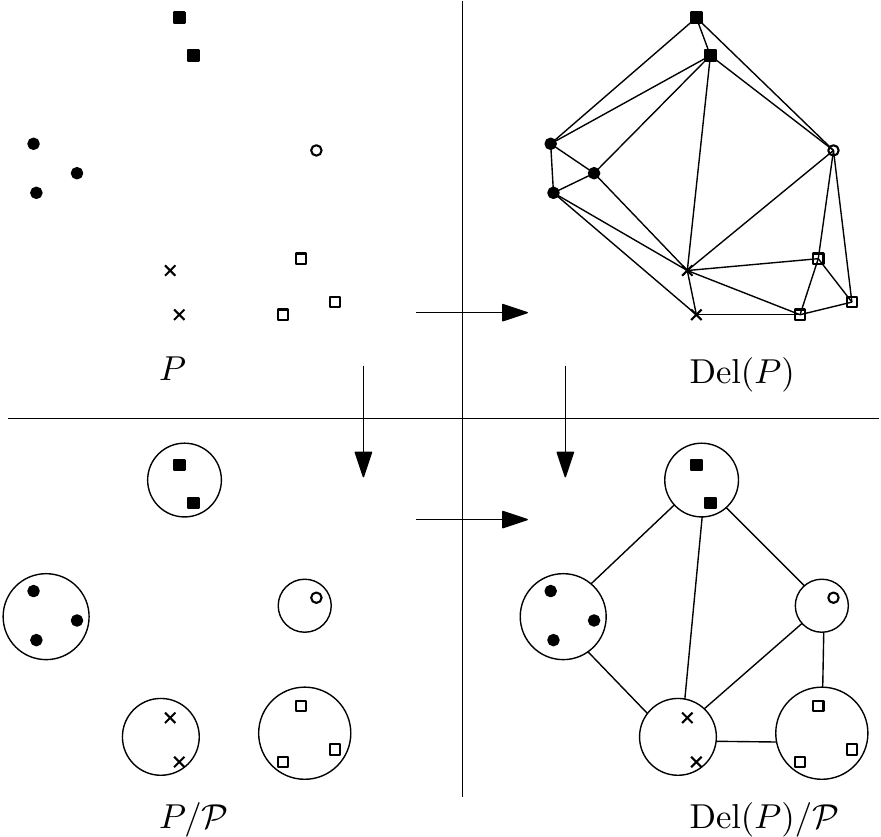}
\caption{Illustration of Lemma~\ref{thm-quotient}.}
\end{figure}

\begin{lemma}\label{thm-quotient}Let $G=(V,E)$
be a complete geometric graph, $\mathcal{V}$ be a partition of $V$ and $S$ be a $t$-spanner of $G$.
Then $S/\mathcal{V}$ is a $t$-spanner of $G/\mathcal{V}$.
\end{lemma}
\old{
\begin{proof} Let $(U,W)$ be an edge of $G/\mathcal{V}$ and $(u,w)$
be an edge of $G$ such that $|uw|=|UW|$. Since $G$ is complete, the edge $(u,w)$ is in $G$, and
since $S$ is a $t$-spanner of $G$, there is a path $\psi=u_1,\ldots,u_k$ in $S$ such that
$u_1=u,u_k=w$ and the length of $\psi$ is at most $t|uw|$. By definition, $S/\mathcal{V}$ contains
a path $\Psi=U_1,\ldots,U_k$ such that $U_1=U$ and $U_k=W$. The length of $\Psi$ is at most
$$
\sum_{i=1}^{k-1}|U_iU_{i+1}|\leq \sum_{i=1}^{k-1}|u_iu_{i+1}| \leq t|uw| = t|UW|
$$
which means that $\Psi$ is a $t$-spanning path for $(U,W)$ in $S/\mathcal{V}$.
\end{proof}
}
\begin{proof} Let $(U,W)$ be an edge of $G/\mathcal{V}$ and $(u,w)$
be an edge of $G$ such that $|uw|=|UW|$. Since $G$ is complete, the edge $(u,w)$ is in $G$, and
since $S$ is a $t$-spanner of $G$, there is a path $\psi=u_1,\ldots,u_k$ in $S$ such that
$u_1=u,u_k=w$ and the length of $\psi$ is at most $t|uw|$.  For each $u_i$ of $\psi$, let
$U_i\in\mathcal{V}$ be such that $u_i\in U_i$. Notice that it is possible that $U_i=U_{i+1}$ for
some $i$. Let $\Psi$ be the subsequence of $U=U_1,\ldots,U_k=W$ that consists in those $U_i$ such
that $i<k$ and $U_i\neq U_{i+1}$. By definition, the sequence $\Psi$ is a path in $S/\mathcal{V}$
and it consists of at most $k'\leq k$ nodes. The length of $\Psi$ is at most
$$
\sum_{i=1}^{k'-1}|U_iU_{i+1}|\leq \sum_{i=1}^{k-1}|u_iu_{i+1}| \leq t|uw| = t|UW|
$$
which means that $\Psi$ is a $t$-spanning path for $(U,W)$ in $S/\mathcal{V}$.
\end{proof}

\section{The Additively Weighted Delaunay Graph}\label{section-disk-del-spanning-ratio}

\citet{drysdale81} studied a variant of the Voronoi diagram called the Additively Weighted Voronoi
diagram, which is defined as follows: Let $P$ be a weighted point set. The \emph{Additively
Weighted Voronoi diagram} of $P$ is a partition of the plane into $|P|$ regions such that each
region contains exactly the points in the plane having the same closest neighbor in $P$ according
to the additive distance. In other words, the Voronoi cell of a pair $(p_i,r_i)$ contains the
points $p$ such that $d(p,p_i)$ is minimum over all other pairs in $P$. The \emph{Additively
Weighted Delaunay graph} (AW-Delaunay graph) is defined as the face-dual of the Additively Weighted
Voronoi diagram.

Alternatively, if all $r_i$ are positive and for all $i,j$, we have $|p_ip_j|\geq r_i+r_j$, then
the pairs $(p_i,r_i)$ can be seen as disks $D_i$ of radius $r_i$ centered at $p_i$ and $d(p,D_i)$
is the minimum $|pq|$ over all $q\in D_i$. For a set $\mathcal{D}$ of disks in the plane, we
denote the AW-Delaunay graph computed from $\mathcal{D}$ as $\DEL(\mathcal{D})$. When no two disks
intersect, the AW-Delaunay graph is a natural generalization of the Delaunay graph of a set of
points. We say that two disks $A$ and $B$ \emph{properly} intersect if $|A\cap B|>1$.

\begin{proposition}\label{prop-dual} Let $\mathcal{D}$ be a set of disjoint disks in the
plane, and $A,B\in\mathcal{D}$. The edge $(A,B)$ is in $\DEL(\mathcal{D})$ if and only if there is
a disk $C$ that is tangent to both $A$ and $B$ and does not properly intersect any other disk in
$\mathcal{D}$.
\end{proposition}
\begin{proof}Suppose $(A,B)$ is in $\DEL(\mathcal{D})$, and let $c$
be a point on the boundary of the Voronoi cells of $A$ and $B$ and $r$ be the distance from $c$ to
$A$. Since $c$ is equidistant from $A$ and $B$, it is also at distance $r$ from $B$. This means that
the disk $C$ centered at $c$ is tangent to both $A$ and $B$. This disk cannot properly intersect
any other disk of $\mathcal{D}$, since this would contradict the fact that $c$ is in the Voronoi
cells of $A$ and $B$. Similarly, if there is a disk that is tangent to both $A$ and $B$ but does
not properly intersect any other disk of $\mathcal{D}$, then $A$ and $B$ are Voronoi neighbors.
\end{proof}

\begin{figure}
\centering
\includegraphics{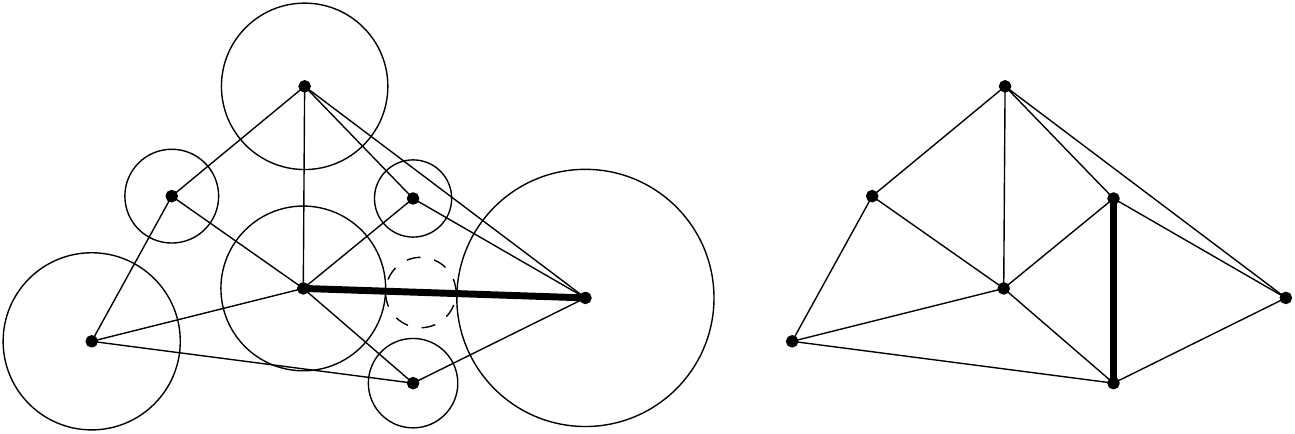}
\caption{The Additively Weighted Delaunay graph compared with the Delaunay graph of the disks
centers.}\label{fig-del-disks}
\end{figure}

Note that the Additively Weighted Delaunay graph is not necessarily isomorphic
to the Delaunay graph of the centers of the disks (see
Figure~\ref{fig-del-disks}). When all radii are equal, however, the two graphs
coincide. We now show that if $\mathcal{D}$ is a set of disks in the plane,
then $\DEL(\mathcal{D})$ is a spanner of $\mathcal{D}$. The intuition behind
the proof is the following: we show the existence of a finite set of points $P$
such that $K(P)/\mathcal{D}$ (where $K(P)$ is the complete graph with vertex
set $P$) is isomorphic to the complete graph on $\mathcal{D}$ and
$\DEL(P)/\mathcal{D}$ is a subgraph of $\DEL(\mathcal{D})$. Then, we use
Lemma~\ref{thm-quotient} to prove that $\DEL(P)/\mathcal{D}$ is a spanner of
$\mathcal{D}$, which implies that $\DEL(\mathcal{D})$ is a spanner of
$\mathcal{D}$.

\begin{definition}\label{def-repr} Let $A,B$ be disjoint disks and $S$ a set of points such
that $A\cap S=\emptyset$ and $B\cap S=\emptyset$. A set of points $R$
\emph{represents} $S$ with respect to $A$ and $B$ if for every disk $F$ that is tangent to both $A$ and
$B$, we have $F\cap S\neq\emptyset\Rightarrow F\cap R\neq\emptyset$. If $\mathcal{D}$ is a set of
disjoint disks, then a set of points $\mathcal{R}$ \emph{represents} $\mathcal{D}$ if for all
$A,B,C\in\mathcal{D}$, there is a subset of $\mathcal{R}$ that represents $C$ with respect to $A$
and $B$.
\end{definition}
From here to the end of the proof of Lemma~\ref{lemma-repr}, unless stated otherwise, let
\begin{enumerate}
\item $A,B$ be two disjoint disks in the plane having their
center on the $x$-axis;
\item $D(y)$ be the disk that is tangent to both $A$ and $B$ and whose center has
$y$-coordinate equal to $y$;
\item $y(D)$ be the $y$-coordinate of the center of a disk $D$;
\item $\ell_1,\ell_2$ be the two lines that are outer-tangent to both $A$ and $B$ (respectively, from below and above);
\item $y_1,y_2$ be such that $y_1<y_2$ and $D(y_1)\cap D(y_2)\neq \emptyset$;
\item $\ell$ be the line through the intersection points of the boundaries of $D(y_1)$ and $D(y_2)$
(if $D(y_1)$ and $D(y_2)$ are tangent, then $\ell$ is the unique line that is tangent to both
$D(y_1)$ and $D(y_2)$);
\item $T(A,B)$ denote the region below $\ell_2$, above $\ell_1$ and between
$A$ and $B$; and
\item $l^+$ ($l^-$) be the closed half-plane above (below) a non-vertical line $l$.
\end{enumerate}
Throughout this section, it is implicitly assumed that $D(\infty)=\ell_2^+$ and $D(-\infty)=\ell_1^-$.
\begin{figure}
\centering\includegraphics{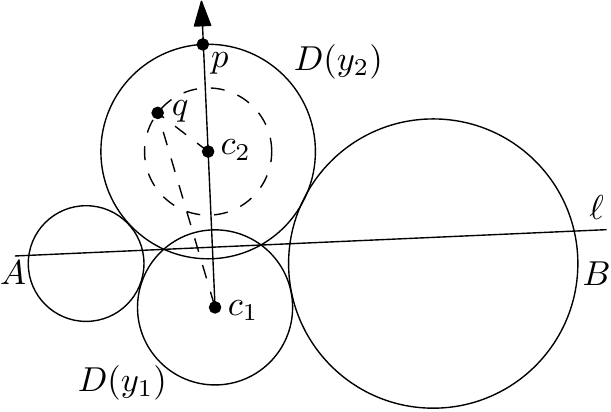}\caption{Illustration of the proof of
Lemma~\ref{lemma-lunes}.}\label{fig-lemma-lunes}
\end{figure}

\begin{lemma}\label{lemma-lunes}  Given $y_1<y_2$ and $D(y_1)\cap D(y_2)\neq \emptyset$, we have $D(y_1)\cap\ell^+\subset D(y_2)\cap\ell^+$ and
$D(y_2)\cap\ell^-\subset D(y_1)\cap\ell^-$ (see Figure~\ref{fig-lemma-lunes}).
\end{lemma}
\begin{proof}
Notice that either $D(y_1)\cap\ell^+\subset D(y_2)\cap\ell^+$ or $D(y_2)\cap\ell^+\subset
D(y_1)\cap\ell^+$. Therefore, all we need to show is that $(D(y_2)\cap\ell^+)\setminus
(D(y_1)\cap\ell^+)$ is not empty. Let $c_1,c_2$ be the respective centers of $D(y_1)$ and $D(y_2)$,
and $p$ be the intersection point of the infinite ray from $c_1$ through $c_2$ with the boundary of
$D(y_1)\cup D(y_2)$.

We show by contradiction that $p$ is not in $D(y_1)$. If that was the case, then $D(y_2)$ would be
completely contained in $D(y_1)$. The reason for this is that there is no point of $D(y_2)$ that is
farther from $c_1$ than $p$. Let $q$ be a point of $D(y_2)$. Then $|qc_1|\leq |qc_2|+|c_2c_1|\leq
|pc_2|+|c_2c_1|=|pc_1|$. But the fact that $D(y_2)$ is completely contained in $D(y_1)$ contradicts
the fact that they are both tangent to $A$ and $B$.

Therefore, since $p\in \ell^+$, we have $p\in (D(y_2)\cap\ell^+)\setminus (D(y_1)\cap\ell^+)$,
which imply that $D(y_1)\cap\ell^+\subset D(y_2)\cap\ell^+$. Similarly, $D(y_2)\cap\ell^-\subset
D(y_1)\cap\ell^-$.
\end{proof}

\begin{figure}
\centering
\includegraphics{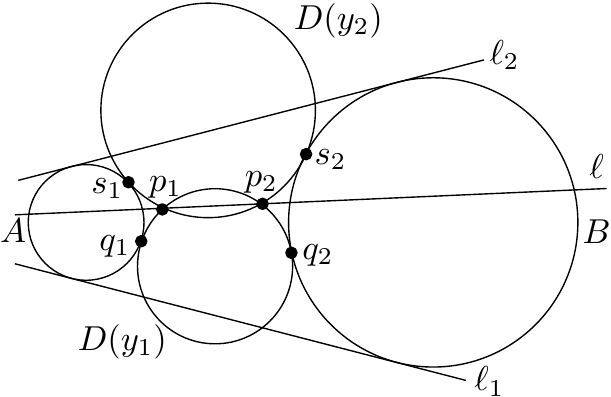}\caption{Illustration of the proof of Lemma~\ref{lemma-tunnel}.}\label{fig-lemma-tunnel}
\end{figure}

\begin{lemma}\label{lemma-tunnel}
Let $p_1,p_2$ be the intersection points of the boundaries of $D(y_1)$ and $D(y_2)$ (if $D(y_1)$
and $D(y_2)$ are tangent, then $p_1=p_2$). Then $p_1$ and $p_2$ are in $\ell_2^-$ and in $\ell_1^+$
(see Figure~\ref{fig-lemma-tunnel}).
\end{lemma}
\begin{proof}
Let $q_1,q_2$ be the tangency points of $D(y_1)$ with $A$ and $B$ and $s_1,s_2$ be the tangency
points of $D(y_2)$ with $A$ and $B$. By Lemma~\ref{lemma-lunes}, $q_1,q_2$ are below $\ell$ and
$s_1,s_2$ are above $\ell$. Since $\ell$ is above $q_1$ and $q_2$, which are in turn above
$\ell_1$, it follows that $p_1$ and $p_2$ are above $\ell_1$. By a symmetric argument, $p_1$ and
$p_2$ are below $\ell_2$.
\end{proof}

\begin{lemma}\label{lemma-containment} The following are true:
\begin{enumerate}
\item For all $p\in\ell_2^+$, there exists a line $y=y_0$ such that for all disk $E$ that is tangent to both $A$ and
$B$, if the center of $E$ is above $y_0$ then $p\in E$.
\item For all $p\in\ell_1^-$, there exists a line $y=y_1$ such that for all disk $E$ that is tangent to both $A$ and
$B$, if the center of $E$ is below $y_1$ then $p\in E$.
\item For all $p$ in $T(A,B)$, there exists two lines $y=y_0$ and $y=y_1$ such that for all disk $E$ that is tangent to both $A$ and
$B$, $p\in E$ if and only if the center of $E$ is between $y_0$ and $y_1$.
\end{enumerate}
\end{lemma}
\begin{proof}
For (1), the existence of $y_0$ is guaranteed by the fact that $\lim_{y\rightarrow\infty}
D(y)=\ell_2^+$. Now, let $y_0$ be such that $p\in D(y_0)$ and $y'>y_0$. Let $L(y_0)$ and $L(y')$ be
the lunes respectively defined by the intersection of $D(y_0)$ and $D(y')$ with the half-plane
above $\ell_2$. By Lemma~\ref{lemma-tunnel}, the two points where the boundaries of $D(y_0)$ and
$D(y')$ intersect are below $\ell_2$. Therefore, we have either $L(y_0)\subset L(y')$ or
$L(y')\subset L(y_0)$. But since $y'>y_0$, by Lemma~\ref{lemma-lunes} we have $L(y_0)\subset L(y')$
and therefore $p\in L(y')$. The proof of (2) is symmetric.

\begin{figure}
\centering\includegraphics{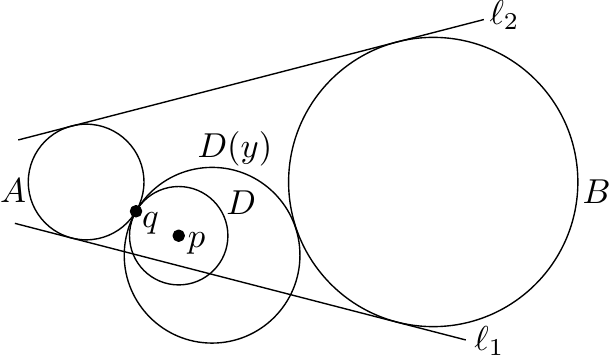}\caption{Illustration of the proof of
Lemma~\ref{lemma-containment} (3) (first part).} \label{fig-lemma-cont-tunnelb}
\end{figure}

\begin{figure}
\centering\includegraphics{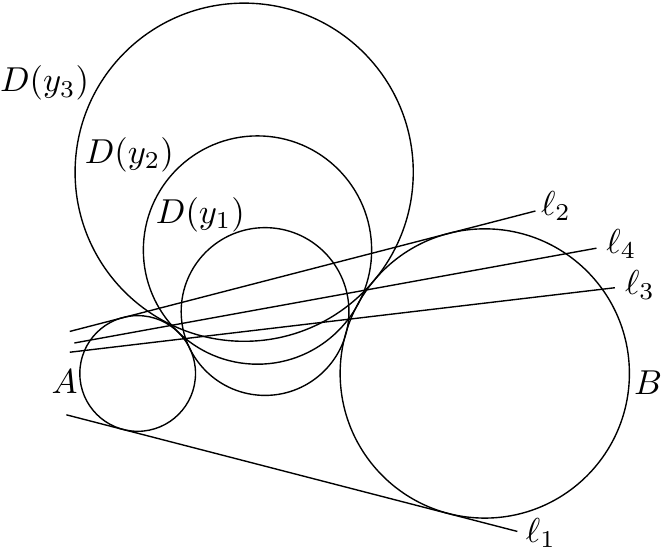}\caption{Illustration of the proof of
Lemma~\ref{lemma-containment} (3) (second part).} \label{fig-lemma-cont-tunnel}
\end{figure}

For (3), the existence is easy to show. Without loss of generality, assume
$d(p,A) \leq d(p,B)$.  Let $D$ be the disk centered at $p$ that is tangent to
$A$ and let $q$ be the tangency point of $A$ and $D$ see
Figure~\ref{fig-lemma-cont-tunnelb}. Since $q\in T(A,B)$, there exists $y$ such
that $D(y)\cap A=q$. Since $D\subseteq D(y)$, there exists a disk that is tangent
to both $A$ and $B$ and contains $p$.

%
%
We now show that $y_1<y_2<y_3$ implies $D(y_1)\cap D(y_3)\subseteq D(y_2)$ (see
Figure~\ref{fig-lemma-cont-tunnel}). Let $\ell_3$ be the line through the intersection points of
the boundaries of $D(y_1)$ and $D(y_2)$ and let $\ell_4$ be the line through the intersection
points of the boundaries of $D(y_2)$ and $D(y_3)$. Let $p\in D(y_1)\cap D(y_3)$. Since $\ell_4$ is
above $\ell_3$ in $D(y_1)\cap D(y_3)$, $p$ is either above $\ell_3$, below $\ell_4$ or both. If
$p\in\ell_3^+$, then since $y_1<y_2$, by Lemma~\ref{lemma-lunes} we have that
$D(y_1)\cap\ell_3^+\subseteq D(y_2)\cap\ell_3^+$ and $p\in D(y_1)\cap D(y_2)$. Similarly, if
$p\in\ell_4^-$, then since $y_2<y_3$, by Lemma~\ref{lemma-lunes} we have that
$D(y_3)\cap\ell_4^-\subseteq D(y_2)\cap\ell_4^-$ and $p\in D(y_3)\cap D(y_2)$. In either case,
$p\in D(y_2)$, which completes the proof.
\end{proof}
\begin{figure}
\centering\includegraphics{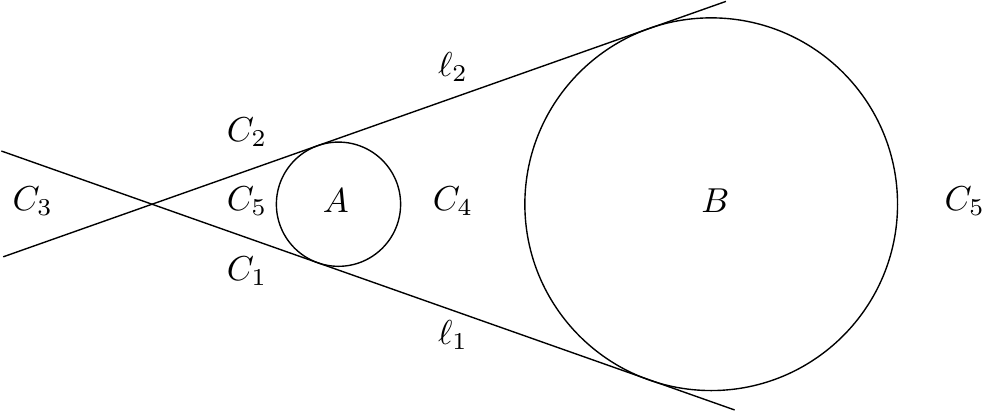}\caption{The five regions for
Lemma~\ref{lemma-repr}.}\label{fig-lemma-repr}
\end{figure}
\begin{figure}
\centering\includegraphics{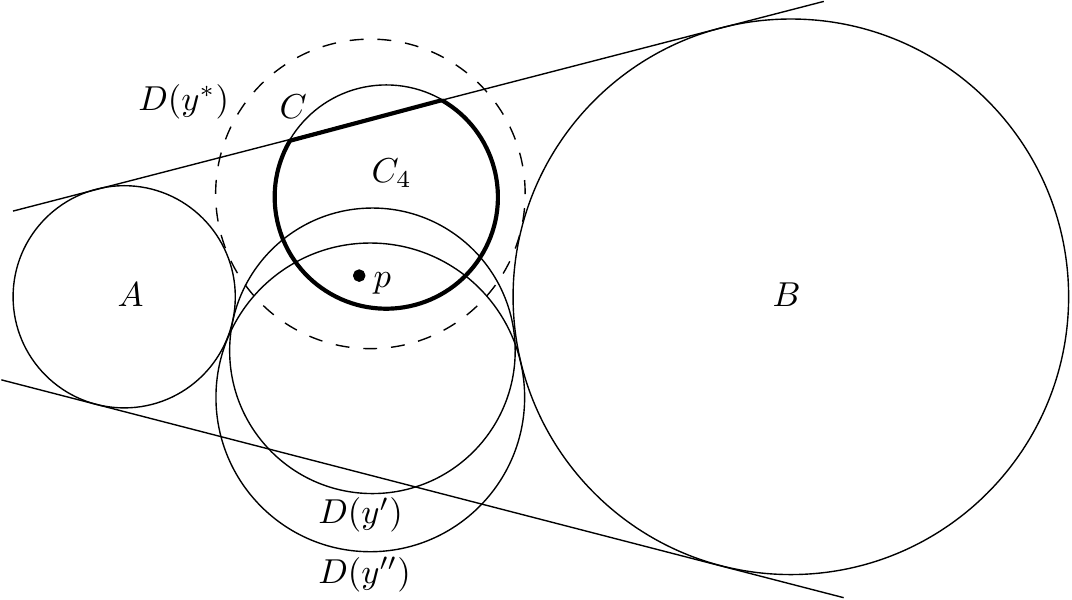}\caption{Case $C_4$ of the proof of
Lemma~\ref{lemma-repr}.}\label{fig-lemma-repr4}
\end{figure}
\begin{lemma}\label{lemma-repr} Let $C$ be a disk that is disjoint of both $A$ and $B$. There exists a
set of at most six points that represents $C$ with respect to $A$ and $B$.
\end{lemma}
\begin{proof} Let
\begin{eqnarray*}
C_1 & := & (C\cap\ell_1^-)\setminus\ell_2^+\\
C_2 & := & (C\cap\ell_2^+)\setminus\ell_1^-\\
C_3 & := & C\cap\ell_1^-\cap\ell_2^+\\
C_4 & := & C\cap T(A,B)\\
C_5 & := & (C\cap\ell_1^+\cap\ell_2^-)\setminus T(A,B)
\end{eqnarray*}
These five regions partition the disk $C$ (see Figure~\ref{fig-lemma-repr}). We show that for each
region, there is a finite set of points that represents it. The cardinality of the union of the
sets is no more than six.

If $C_1\neq\emptyset$, then let $y_0$ be the minimum $y$ such that $D(y)$ intersects $C_1$. Let
$p_1\in C_1\cap D(y_0)$. By definition of $y_0$, for any disk $E$ that is tangent to both $A$ and $B$
and intersects $C_1$, we have $y(E)\geq y_0$, and by Lemma~\ref{lemma-containment}, we have $p_1\in
E$.

Similarly, if $C_2\neq\emptyset$, then let $y_1$ be the maximum $y$ such that $D(y)$ intersects
$C_2$. Let $p_2\in C_2\cap D(y_1)$. By definition of $y_1$, for any disk $E$ that is tangent to both $A$
and $B$ and intersects $C_2$, we have $y(E)\leq y_1$, and by Lemma~\ref{lemma-containment}, we have
$p_2\in E$.

If $C_3\neq\emptyset$, then let $y_0$ be the minimum $y>0$ such that $D(y)$ intersects $C_3$ and
$y_1$ as the maximum $y<0$ such that $D(y)$ intersects $C_3$. Let $p_3\in C_3\cap D(y_0)$ and
$p_4\in C_3\cap D(y_1)$. By definition of $y_0$, for any disk $E$ with $y(E)>0$ that is tangent to both
$A$ and $B$ and intersects $C_3$, we have $y(E)\geq y_0$, and by Lemma~\ref{lemma-containment}, we
have $p_3\in E$. The same reasoning applies to $p_4$ when $y(E)<0$.

If $C_4\neq\emptyset$, then let $y_0$ be the minimum $y$ such that $D(y)$ intersects $C_4$ and
$y_1$ as the maximum $y$ such that $D(y)$ intersects $C_4$. Let $p_5\in C_4\cap D(y_0)$ and $p_6\in
C_4\cap D(y_1)$. Let $y^*$ be such that $C\subseteq D(y^*)$ (see Figure~\ref{fig-lemma-repr4}).
Let $E$ be a disk that is tangent to both $A$ and
$B$ and intersects $C_4$. We show that $y(E)\leq y^*\implies p_5\in E$ (and similarly, $y(E)\geq
y^*\implies p_6\in E$). It is sufficient to show that $y''<y'<y^*\implies C\cap D(y'')\subset C\cap
D(y')$. Let $p\in D(y'')\cap C$. By Lemma~\ref{lemma-containment}, $\exists y_0(p),y_1(p)$ such
that $\forall$ disk $E$ tangent to both $A$ and $B$, we have $y_0(p)\leq y(E)\leq y_1(p)\Leftrightarrow
p\in E$. Therefore, the following hold:
\begin{eqnarray*}
y_0(p)\leq y^* &\leq & y_1(p)\\
y_0(p)\leq y'' &\leq & y_1(p)
\end{eqnarray*}
But since $y''<y'<y^*$, we have $y''<y'<y^*$, which imply that $p\in
 C\cap D(y')$.

Finally, since $C_5\cap E=\emptyset$ for any disk $E$ that is tangent to both $A$ and $B$, there is no
need to select representative points for $C_5$.
\end{proof}

Careful analysis of the proof of Lemma~\ref{lemma-repr} allows us to observe that in fact, only two
points are necessary to represent a disk $C$ with respect to two other disks $A$ and $B$. First,
note that $C_4\neq\emptyset\implies C_3=\emptyset$ and $C_3\neq\emptyset\implies C_4=\emptyset$.
This reduces to four the number of points that are necessary. Also, if $C_1\neq\emptyset$ and
$C_4\neq\emptyset$, then $p_6$ is on $\ell_2$ and is not required since any disk that contains it
also intersects $C_1$ and therefore contains $p_1$. Similarly, if $C_2\neq\emptyset$ and
$C_4\neq\emptyset$, then $p_5$ is not required since any disk that contains it also intersects
$C_2$ and therefore contains $p_2$. Therefore, if $C_4\neq\emptyset$, then the number of points
that are necessary is at most two. A similar argument applies to the case where $C_3\neq\emptyset$.
Finally, if both $C_3$ and $C_4$ are empty, then only $p_1$ and $p_2$ may be required. Therefore,
we have the following corollary:

\begin{corollary}\label{cor-finite-rep} Let $\mathcal{D}$ be a set of $n$ disjoint disks.
There exists a set of at most $2\binom{n}{3}$ points that represents $\mathcal{D}$.
\end{corollary}

\begin{figure}
\centering
\includegraphics{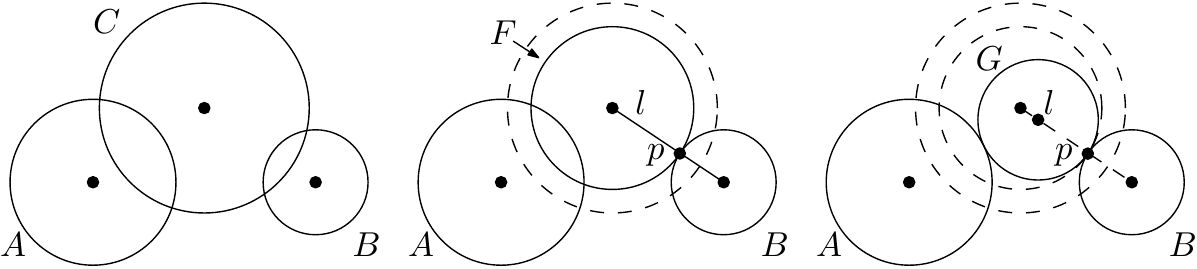}\caption{Proof of Lemma~\ref{lemma-witness-constr}.}\label{fig-witness-constr}
\end{figure}

\begin{lemma}\label{lemma-witness-constr} Let $A$ and $B$ be two disjoint disks and $C$ be a
disk intersecting both of them. Then there exists a disk $G$ inside $C$ that is tangent to both $A$
and $B$.
\end{lemma}
\begin{proof} We show how to construct $G$. Let $a,b,c$ and $r_A,r_B,r_C$ respectively be the
centers and radii of $A,B$ and $C$. Without loss of generality, assume $|ac|-r_C\leq |bc|-r_B$. Let
$F$ be the disk centered at $c$ and having radius $r_F=|bc|-r_B$ (see
Figure~\ref{fig-witness-constr}). The disk $F$ is tangent to $B$. If $F$ is also tangent to $A$,
then let $G=F$ and we are done. Otherwise, $F$ is properly intersecting $A$. In that case, let $p$
be the tangency point of $F$ and $B$, $l$ be the line through $b$ and $c$, and $G$ be the disk
through $p$ having its center on $l$ and tangent to $A$. The result follows from the fact that $G$
is tangent to $B$ and inside $C$.
\end{proof}

\begin{figure}
\centering
\includegraphics{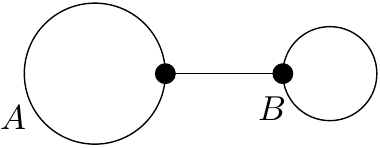}\caption{The distance points of $A$ and
$B$.}\label{fig-dist-points}
\end{figure}

\begin{definition} Let $A$ and $B$ be two disks in the plane. The
\emph{distance points} of $A$ and $B$ are the two ends of the shortest line segment between $A$ and
$B$ (see Figure~\ref{fig-dist-points}). If $\mathcal{D}$ is a set of disjoint disks, then the set
of \emph{distance points} of $\mathcal{D}$ is the set containing the distance points of every pair
of disks in $\mathcal{D}$.
\end{definition}

\begin{figure}
\centering
\includegraphics{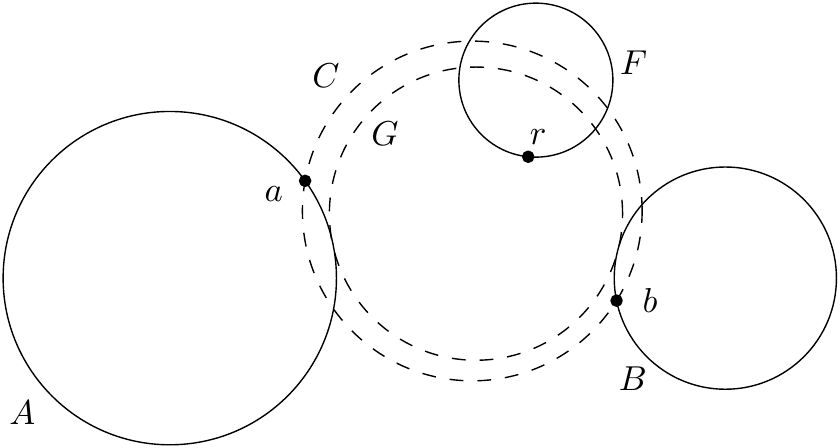}
\caption{Illustration of the proof of Theorem~\ref{thm-del-disks}.}
\end{figure}

\begin{theorem}\label{thm-del-disks} Let $\mathcal{D}$ be a set of $n$ disjoint disks.
Then $\DEL(\mathcal{D})$ is a $t$-spanner of $\mathcal{D}$, where $t$ is the spanning ratio of the
Delaunay triangulation of a set of points.
\end{theorem}
\begin{proof}
By Corollary~\ref{cor-finite-rep}, let $R$ be a set of size at most $2\binom{n}{3}$ that represents
$\mathcal{D}$, let $S$ be the set of distance points of $\mathcal{D}$, and let $P=R\cup S$. Since
$\DEL(P)$ is a $t$-spanner of $P$, by Lemma~\ref{thm-quotient}, we have $\DEL(P)/\mathcal{D}$ is a
$t$-spanner of $K(P)/\mathcal{D}$, where $K(P)$ is the complete graph with vertex set $P$.
Since $P$ contains the distance points of $\mathcal{D}$,
$K(P)/\mathcal{D}$ is isomorphic to the complete graph defined on $\mathcal{D}$. We show that each
edge $(A,B)$ of $\DEL(P)/\mathcal{D}$ is in $\DEL(\mathcal{D})$. Let $(A,B)$ be an edge of
$\DEL(P)/\mathcal{D}$. This means that in $P$, there are two points $a$ and $b$ with $a\in A,b\in
B$ such that there is an empty circle $C$ through $a$ and $b$. By Lemma~\ref{lemma-witness-constr},
$C$ contains a disk $G$ that is tangent to both $A$ and $B$. The disk $G$ is a witness of the
presence of the edge $(A,B)$ in $\DEL(\mathcal{D})$. If that was not the case, this would mean that
there exists a disk $F\in\mathcal{D}$ such that $G\cap F\neq\emptyset$. By definition of $R$, this
implies that $G\cap R\neq\emptyset$ and thus $C\cap P\neq\emptyset$, which contradicts the fact
that $C$ is an empty circle. Therefore, the edge $(A,B)$ is in $\DEL(\mathcal{D})$.
Since $\DEL(P)/\mathcal{D}$ is a $t$-spanner of $\mathcal{D}$ and a subgraph of
$\DEL(\mathcal{D})$, we conclude that $\DEL(\mathcal{D})$ is a $t$-spanner of $\mathcal{D}$.
\end{proof}


\begin{figure}
\centering\includegraphics{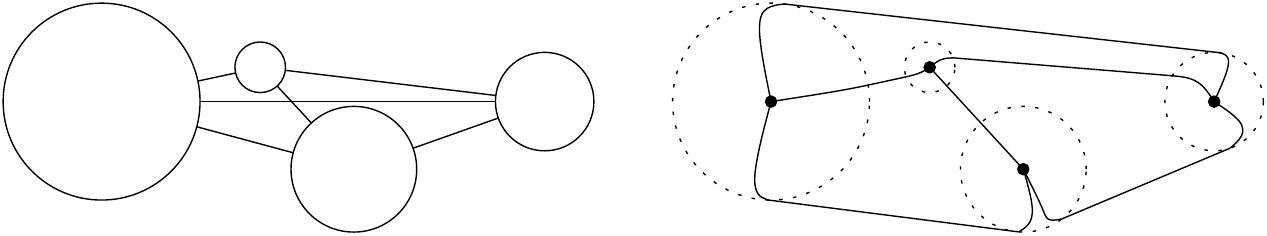}\caption{Even if the embedding of the AW-Delaunay graph
that consists of straight line segments between the centers of the disks is not necessarily a plane
graph, it is planar.}\label{fig-not-plane}
\end{figure}

Note that the embedding of the AW-Delaunay graph that consists of straight line segments between
the centers of the disks is not necessarily a plane graph (see Figure~\ref{fig-not-plane}).
However, the Voronoi diagram of a set of disks $\mathcal{D}$, denoted $\VOR(\mathcal{D})$, is
planar~\cite{okabe00}. Since $\DEL(\mathcal{D})$ is the face-dual of $\VOR(\mathcal{D})$, it is
also planar. An important characteristic of the Delaunay graph of a set of points regarded as a
spanner is that it is a plane graph. Therefore, a natural question is whether $\DEL(\mathcal{D})$
has a plane embedding that is also a spanner.

The proof of Theorem~\ref{thm-del-disks} suggests the existence of an algorithm allowing to compute
such an embedding: compute the Delaunay triangulation of the set $P$ that contains the distance
points and the representative of $\mathcal{D}$. The graph $\DEL(P)$ can be regarded as a multigraph
whose vertex set is $\mathcal{D}$. Then, for each pair of disks that share one or more edges, just
keep the shortest of those edges. This simple algorithm allows to compute a plane embedding of
$\DEL(\mathcal{D})$ that is also a spanner of $\mathcal{D}$. However, its running time is
$O(n^3\log n)$. Whether or not it is possible to compute a plane embedding of $\DEL(\mathcal{D})$
that is also a spanner of $\mathcal{D}$ in a better running time remains a open question.

\old{

In this section, we show that $\DEL(\mathcal{D})$ has a plane embedding that has the same spanning
ratio as the Delaunay graph of a set of points and that given $\DEL(\mathcal{D})$ (which can be
computed in time $O(n\log n)$), this embedding can be computed in time $O(n)$.

\begin{definition} Let $\mathcal{D}$ be a finite set of disjoint disks, and $A,B\in \mathcal{D}$
such that the edge $(A,B)$ is in $\DEL(\mathcal{D})$. The points $p_1,p_2$ are \emph{witnesses} of
the edge $(A,B)$ if there is a disk $F$ such that
\begin{enumerate}
\item $p_1\in(A\cap F)$ and $p_2\in(B\cap F)$ and
\item $F$ does not intersect any other disk of $\mathcal{D}$.
\end{enumerate}
They are also \emph{minimum witnesses} if the distance between them is minimum. Notice that in that
case, the disk $F$ has minimum radius and is tangent to $A$ and $B$. A set of points $W$ is a
\emph{(minimum) witness} of $\DEL(\mathcal{P})$ if it contains (minimum) witnesses for all edges in
$\DEL(\mathcal{D})$.
\end{definition}

\begin{lemma} Let $\mathcal{D}$ be a finite set of disjoint disks. Given $\DEL(\mathcal{D})$,
a minimum witness set of $\DEL(\mathcal{D})$ can be computed in time $O(n)$.
\end{lemma}
\begin{proof}
We show that given $\VOR(\mathcal{D})$, minimum witnesses can be found for each edge of
$\DEL(\mathcal{D})$ in amortized constant time. Let $A,B\in \mathcal{D}$ such that the edge $(A,B)$
is in $\DEL(\mathcal{D})$. Notice that any disk that is tangent to both $A$ and $B$ has its center
on the bisector $b$ of $A$ and $B$. Let $D$ be the disk that is tangent to both $A$ and $B$ whose
center $d$ is the intersection of $b$ with the line through the centers of $A$ and $B$. If $d$ is
on the common boundary of the Voronoi regions of $A$ and $B$, then since $D$ has minimum radius
among all disks that are tangent to both $A$ and $B$, the tangency points of $D$ with $A$ and $B$
are minimum witnesses of the edge $(A,B)$. Otherwise, let $V$ be the vertices defining the common
boundaries of the Voronoi regions of $A$ and $B$ (note that all points of $V$ are on $b$). The
center of the disk defining the minimum witnesses of $(A,B)$ is in $V$. Since $\VOR(\mathcal{D})$
has linear complexity~\cite{fortune87}, $V$ has constant size on average. The result follows from
the fact that since $\DEL(\mathcal{D})$ is planar, it has a linear number of edges.
\end{proof}

\begin{theorem}\label{reverse-prop} Let $\mathcal{D}$ be a finite set of disjoint disks.
Then $\DEL(\mathcal{D})$ has a planar embedding that is a $t$-spanner of $\mathcal{D}$, where $t$
is the spanning ratio of the Delaunay triangulation of a set of points. Moreover, given
$\DEL(\mathcal{D})$, this embedding can be computed in time $O(n)$.
\end{theorem}
\begin{proof} The algorithm proceeds as follows: for every edge $(D_1,D_2)$ of $\DEL(\mathcal{D})$,
draw an edge $(w_1,w_2)$ where $w_1,w_2$ form a pair of minimum witnesses of the edge $(D_1,D_2)$.
Let $W$ be the set of all minimum witnesses and $G$ the resulting graph. The only difference
between $G$ and $\DEL(\mathcal{D})$ is the length of the edges.

To show that $G$ is a $t$-spanner of $\mathcal{D}$, consider the set $P$ that is the union of a
finite set that represents $\mathcal{D}$ and the distance points of $\mathcal{D}$. Let
$D_1,D_2\in\mathcal{D}$ and $p_1,p_2$ be the distance points of $D_1$ and $D_2$. In $\DEL(P)$,
there is a $t$-spanning path $\rho$ from $p_1$ to $p_2$. The existence of this path implies the
existence of a $t$-spanning path in $G$ from $D_1$ to $D_2$: Let $e$ be an edge of $\rho$. We show
that $G$ contains an edge whose adjacent vertices are on the same disks as $e$ and whose length is
at most $|e|$. The only case that we need to consider is when the two endpoints of $e$ belong to
two different disks $F_1$ and $F_2$. In that case, from the proof of Theorem~\ref{thm-del-disks},
the edge $(F_1,F_2)$ is in $\DEL(\mathcal{D})$ and the endpoints of $e$ are witnesses of that edge.
By definition of $G$, it contains an edge $(f_1,f_2)$ where $f_1$ and $f_2$ are minimum witnesses
of the edge $(F_1,F_2)$. Since it is minimum, its length is at most $|e|$, which completes the
proof.
\end{proof}

}

\section{Conclusion}\label{section-disk-del-conclusion}

In this paper, we showed how, given a weighted point set where weights are positive and
$|p_ip_j|\geq r_i+r_j$ for all $i\neq j$, it is possible to compute a $(1+\epsilon)$-spanner of
that point set that has a linear number of edges. We also showed that the Additively Weighted
Delaunay graph is a $t$-spanner of an additively weighted point set in the same case. The constant
$t$ is the same as for the Delaunay triangulation of a point set (the best current value is
2.42~\cite{keil92}). We could not see how the Well-Separated Pair Decomposition (WSPD) can be
adapted to solve that problem. The first difficulty resides in the fact that it is not even clear
that, given a weighted point set, a WSPD of that point set always exists. Other obvious open
questions are whether our results still hold when some weights are negative or $|p_ip_j|< r_i+r_j$
for some $i\neq j$. Also, we did not verify whether our variant of the Yao graph can be computed in
time $O(n\log n)$. Finally, another problem that could be explored is whether it is possible to
compute $t$-spanners for multiplicatively weighted point sets.

\bibliographystyle{myBibliographyStyle}
\bibliography{add-weighted-spanners}

\end{document}